# Spin-Hall effect in the scattering of structured light from plasmonic nanowire

Deepak K. Sharma,[1] Vijay Kumar,[1,*] Adarsh B. Vasista,[1] Shailendra K. Chaubey,[1] G. V. Pavan Kumar[1,2]

[1]*Photonics and Optical Nanoscopy Laboratory, Department of Physics, Indian Institute of Science Education and Research, Pune, India-411008.*

[2]Center for Energy Science, Indian Institute of Science Education and Research, Pune, India-411008.

*Corresponding author: vijaykumarsop@gmail.com

Spin-orbit interactions are subwavelength phenomena which can potentially lead to numerous device related applications in nanophotonics. Here, we report Spin-Hall effect in the forward scattering of Hermite-Gaussian and Gaussian beams from a plasmonic nanowire. Asymmetric scattered radiation distribution was observed for circularly polarized beams. Asymmetry in the scattered radiation distribution changes the sign when the polarization handedness inverts. We found a significant enhancement in the Spin-Hall effect for Hermite-Gaussian beam as compared to Gaussian beam for constant input power. The difference between scattered powers perpendicular to the long axis of the plasmonic nanowire was used to quantify the enhancement. In addition to it, nodal line of HG beam acts as the marker for the Spin-Hall shift. Numerical calculations corroborate experimental observations and suggest that the Spin flow component of Poynting vector associated with the circular polarization is responsible for the Spin-Hall effect and its enhancement.

Angular momentum is an inherent property of electromagnetic fields. Rotating electric field vector of the oscillating transverse electric field gives rise to Spin angular momentum (SAM), and azimuthal phase gradient is responsible for the Orbital angular momentum (OAM). These two quantities are intrinsic in nature. In addition to it, there is an extrinsic OAM due to trajectory of the centroid of light beam. Thus, circular polarization, helical phase front and the trajectory of the light are associated with SAM, intrinsic-OAM and extrinsic-OAM respectively [1,2]. In free space propagation, these three forms of AM are independent and conserved. However, because of Spin-orbit coupling, interconversion between different forms of AM is possible.

Transformation of SAM to intrinsic-OAM are reported in high NA focusing, space-variant subwavelength gratings and liquid crystal devices [3–5]. Intrinsic AM (SAM or intrinsic-OAM) to extrinsic-OAM conversion is called as Hall effect (spin-

Hall or orbital-Hall) and also known as optical Magnus effect [6]. This effect is observed in beam reflection or refraction [7,8], propagation through medium [9–11] or metamaterial [12] and scattering [13–15]. Spin-orbit effects cover a wide class of electromagnetic phenomena such as focusing, reflection, scattering, light propagation and light-matter interactions. Therefore, spin-orbit coupling can be utilized in emerging photonic applications for the generation of structured light, enhanced optical manipulation and controlling optical wave propagation [16].

In this letter, we report preferential radiation direction (preferential scattering) governed by the handedness of circular polarization in the forward scattering of structured light from a plasmonic nanowire as a manifestation of the Spin-Hall effect. In the past, manifestation of spin-orbit coupling is reported in scattering of Gaussian beam from a spherically symmetric particle [13,17–19] except for a few reports on the use of structured light or aspherical geometry [14,20,21]. These effects are weak and one of the important aspects is to enhance them in terms of Spin-Hall shift (separation of orthogonal circular polarization) and signal strength (intensity of orthogonal circular polarization analogous to Hall voltage). General consideration is on the enhancement of Spin-Hall shift [22,23]. Plasmonic structures have been proposed as potential candidate for the enhancement of Spin-Hall signal [24,25]. It is also reported that structured-light matter interaction enhances the optical manipulation [16]. Motivated by this, we use a combination of structured light and a plasmonic nanowire to enhance the Spin-Hall effect. We found that there is significant enhancement in the Spin-Hall signal for Hermite-Gaussian beam over Gaussian beam. The additional advantage of using Hermite-Gaussian beam is that, nodal line acts as the marker of the Spin-Hall shift, i.e. respective orthogonal circular polarization can be found on either side of nodal line.

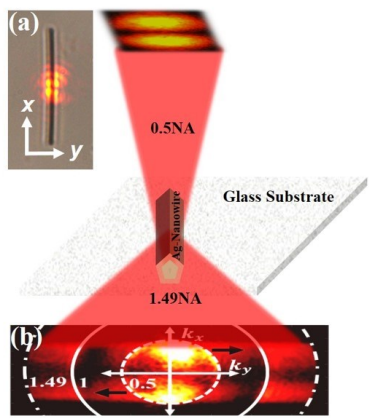

Figure 1. Schematic of the experimental configuration to observe Spin-Hall effect in the scattering from Ag-nanowire. Objective lens of 0.5 NA is used to focus structured light (HG) on the nanowire placed on glass substrate and forward scattered light is collected using 1.49 NA oil immersion object lens. (a) Bright-field image of focused $HG_{10}$ beam on nanowire. (b) Experimentally observed scattered far-field of LCP HG beam. The white circles represent respective NA values and black arrows show the preferential scattering directions.

Experimental observation of the spin-Hall effect was realized using the schematic as shown in Fig. 1. Single-crystalline Ag-nanowires were chemically prepared using a standard protocol [26,27] and drop casted on the glass substrate. Gaussian mode (G) at wavelength 632.8 nm from He-Ne laser was projected on to the spatial light modulator (SLM) with blazed hologram to generate Hermite-Gaussian ($HG_{10}$) beam [28]. Here sub-indices give the number of nodal planes in $x$- and $y$-directions respectively. Polarization of generated beam was controlled using a combination of a polarizer and a waveplate (half or quarter). The polarized beam was focused at the center of the nanowire of diameter ~350nm and few microns in length using an objective lens of 0.5 NA (Fig. 1(a)). The forward scattered light from nanowire was collected using an oil immersion objective lens (1.49 NA). Back focal plane of the collection objective lens was imaged on charged coupled device (CCD) using relay optics (not shown) [29] for Fourier plane (FP) imaging of scattered light (Fig. 1(b)). Figure 1(b) shows a preferential scattering (energy flow) in the direction indicated by black arrows. There are two regions of interest, subcritical or *allowed* (NA<1) and supercritical or *forbidden* (NA>1), depending upon the origin of the light [30]. Manifestation of Spin-Hall effect is prominent within the subcritical region, therefore we have focused only in the region corresponding to 0.5-1.0 NA as shown in Fig. 2. Region corresponding to NA < 0.5 was eliminated for containing unscattered light.

Wire was aligned along the $x$- axis and centered at $y = 0$. Far-field scattered power of left circularly polarized (LCP) HG beam in $+k_y$ and $-k_y$ directions are defined as $P_+(k_x, k_y)$ and $P_-(k_x, k_y)$ respectively, as shown in top and bottom of Fig. 2(a). Subscripts $\pm$ denotes power scattered along $\pm k_y$ directions. Here $k_x = \kappa_x/\kappa_0$ and $k_y = \kappa_y/\kappa_0$ are the normalized wave vector directions and $\kappa_0$ is the free space wave vector. Sum of the scattered power over all $k_y$ is defined as $P_\pm(k_x) = \sum_{k_y} P_\pm(k_x, k_y)$ shown in Fig. 2(b). Fig. 2(d) and (e) show $P(k_x)$ for right circular (RC), left circular and linearly (perpendicular to wire) polarized HG-beam and Gaussian beams respectively. From Fig. 2(d) and (e), it is clear that there is spin controlled preferential radiation direction in the scattered far-field. Black arrows in Fig. 2(c) pictorially show preferential scattering direction for LCP HG beam. To further quantify the amount of preferential scattering, difference of powers $(P_+(k_x) - P_-(k_x))$ was plotted (Fig. 3). It clearly shows that for LCP HG and Gaussian beams, preferential scattering is in $+k_y$ ($-k_y$) direction for $k_x < 0 (> 0)$ forming an antisymmetric scattering pattern, which is reversed on changing the handedness of polarization. It also shows that the amount of preferential scattered power is significantly greater for HG-beam then for the Gaussian beam. It should be noted that powers of HG-beam and Gaussian beam were equal and these beams do not carry any orbital angular momentum. Hence, we have experimentally demonstrated: (1) Spin-Hall shift is on either side of the nodal line at the center of HG-beam, (2) the enhancement of Spin-Hall signal by utilizing structured light and plasmonic nanowire. These aspects of Spin-Hall effect are the result of spin flow [31,32] component of Poynting vector associated with circular polarization, as explained below using simulation results.

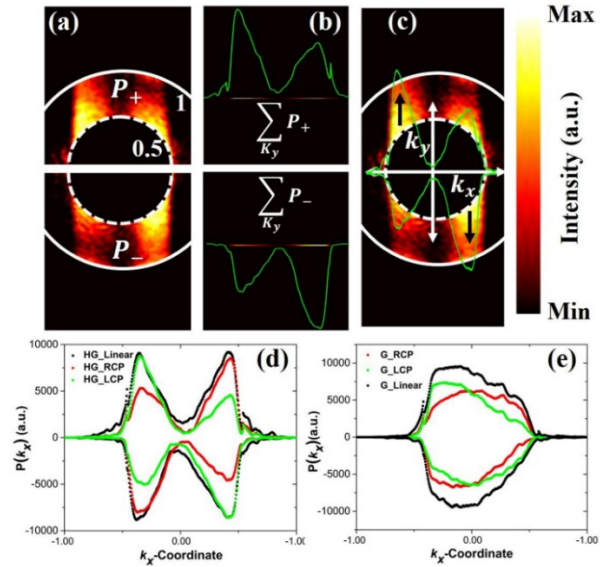

*Figure 2. Analysis of Scattered far-field (Fourier plane image) from Ag-nanowire. (a) Fourier plane image showing region of interest (0.5 < NA <1.0) for observation of Spin Hall effect. (b) Definition of scattered powers along $\pm k_y$ wave vectors i.e. $P_\pm(k_x) = \sum_{k_y} P_\pm(k_x, k_y)$. (c) $P(k_x)$ superimposed on $P(k_x, k_y)$. (d) and (e) Distribution of $P(k_x)$ for LCP (green), RCP (red) and linear (black) Hermite-Gaussian and Gaussian Beams respectively.*

In order to support our experimental results, we carried out the COMSOL Multiphysics 5.1 simulations to calculate the near-field electric field and near-to-far-field transformations were computed by utilizing reciprocity arguments [33]. Ag-nanowire with pentagonal cross-section of diameter $350\ nm$ and $5\mu m$ in length was used in simulations. The wire was placed on glass substrate with refractive index 1.52 and surrounded by air of refractive index 1. Wavelength dependent refractive index of silver was obtained from [34]. Polarized Hermite-Gauss or Gaussian beam of wavelength $633nm$ was focused on to the nanowire-glass interface. The simulation area was meshed with free tetrahedral mesh of minimum element size 10nm and terminated using scattering boundary condition to avoid spurious reflections from the boundaries. Fig. 4 shows the simulation configuration and near-field total Poynting vector in $yz$-plane at $x = 0.1\mu m$ for LCP HG-beam, where black arrows represent the total Poynting vector flow in the plane.

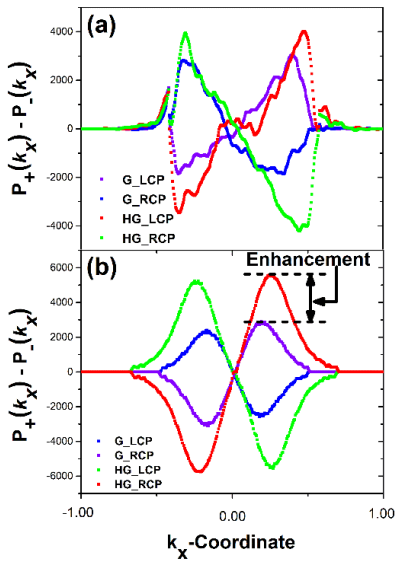

Figure 3. Enhancement of Spin-Hall effect. Difference of power scattered perpendicular to long axis of nanowire, P_+ (k_x )-P_- (k_x). Top (a) and bottom (b) graphs respectively are the experimental and simulation results for Hermite-Gaussian (HG_LCP (green) and HG_RCP (red)) and Gaussian beams (G_LCP (blue) and G_RCP (violet)).

Poynting vector gives the measure of energy flow in optical fields. Total energy flow can be decomposed into 'spin flow' and 'orbital flow' for paraxial fields [31] as well as for nonparaxial fields [32]. For a LCP HG-beam, the calculated spin flow ($\nabla \times Im(\vec{E}^* \times \vec{E})$) is shown in Fig. 4(a). It clearly shows two azimuthally asymmetric spin flow vortices having center coinciding with maximum intensity points ($\pm 0.5 \mu m$) of two lobes of HG-beam. The spin flow is proportional to the transverse gradient of the intensity. Therefore, we chose a $yz$-plane at $x = 0.1 \mu m$ (near maximum spin flow) to demonstrate the effect of spin flow in near- and far-field shown in simulation configuration (Fig. 4). Because of net circulating spin flow in the direction of scattered power along $-k_y$ wavevectors, the total energy flow (shown by black arrows in near field in $yz$ plane at $x = 0.1 \mu m$) makes an angle with respect to normal ($z$-axis) to wire which is possibly giving rise to net energy flow in the $-k_y$ direction in the scattered far-field. Total energy flow will be reversed for $yz$ plane at $x = -0.1 \mu m$ because of opposite spin flow. Hence in the far-field we found a net energy flow (preferential scattering power) in $+k_y(-k_y)$ direction for $k_x < 0 (k_x > 0)$ as shown by black arrows in Fig. 4(b). Similar explanation holds for RCP HG-beam and the corresponding figures are shown Fig. 4(e)-(f). For linearly polarized (perpendicular to the wire) HG-beam there will not be any spin flow, therefore far-field is symmetric with respect to $k_x$-axis without any preferential scattering.

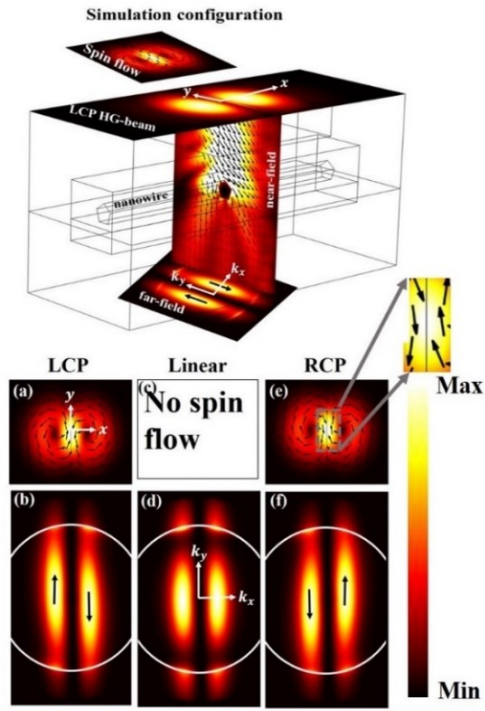

*Figure 4. Simulation configuration and results for HG-beam. (a) and (e) are the spin flow for the LCP and RCP HG-beams. Zoomed in view of center of figure (e) shows zero spin flow along nodal line. Arrows show the local spin flow directions and the background is its magnitude. (b) and (f) are the corresponding scattered far-field respectively. (d) Scattered far-field of HG-beam linearly polarized. Black arrows and white circles in far-field shows the preferential scattering direction and NA value of 1 respectively.*

In contrast to HG-beam, scattered far-field of Gaussian beam has only one intensity lobe instead of two in FP image [35], because of absence of intensity minimum due to no nodal line at center of the beam. Spin flow for LCP and RCP Gaussian beam is given in Fig. 5(a) and (e) respectively. Spin flow vortex has singularity at the center of Gaussian beam with azimuthally symmetric spin flow unlike HG beam. Following the similar line of thought as before, spin flows and corresponding scattered far-field of LCP, linear and RCP Gaussian beams are shown in Fig. 5. Because of vortex spin flow, intensity lob (along $k_y$ axis) in FP image gets weakly rotated in the direction of the handedness of spin flow. Similar kind of energy shift is theoretically reported in the diffraction of circularly polarized Gaussian beam [21]. In addition, directional scattering of circularly polarized light from plasmonic and dielectric scatterers is also reported in [36–38].

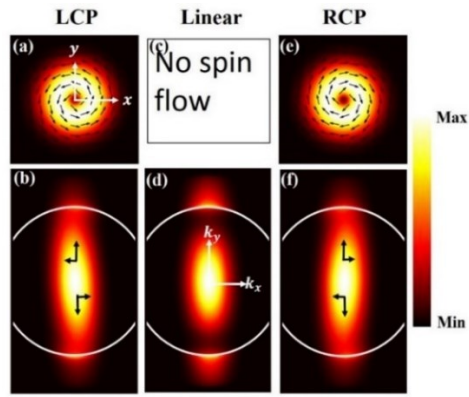

*Figure 5. Simulation results for Gaussian beam. (a), (c) and (e) are the spin flow for the LCP, linear and RCP HG-beams, (b), (d) and (f) are the corresponding scattered far-field respectively. Black arrows in far-fields represents the total energy flow components. Remaining details are same as in Fig. 4.*

Finally, a plot (Fig. 3(b)) of difference of total power scattered ($P_+(k_x) - P_-(k_x)$) perpendicular to the nanowire (as calculated above for experimental data) is extracted from the simulated results. This plot shows the contribution of spin flow on the scattered power from the nanowire. We can see that the maximum preferential scattered power for HG-beam is nearly double as compared to Gaussian beam, thus proves the enhancement of Spin-Hall signal. This enhancement in far-field scattering of HG-beam is because of spin flow component perpendicular to the nanowire is significantly higher than along the nanowire. Whereas for the case of Gaussian beam, spin flow has equal components perpendicular and along the nanowire. Along the nanowire component results in the Spin-Hall shift that is seen as the weak rotation in $\pm k_x$. It is worth noting that, enhancement in the scattered power due to Spin-Hall effect is because of the spin flow component perpendicular to the nanowire. Whereas, Spin-Hall shift arises due to spin flow component along the nanowire.

In summary, we report experimental observation of Spin-Hall effect in the scattering from a plasmonic nanowire. Spin-Hall effect is quantified by taking the difference of scattered powers perpendicular to the long axis of wire and explained on the basis of spin flow component of the Poynting vector associated with circular polarization. Simulation results reveal that scattered power difference is nearly double for HG-beam than for Gaussian beam. With the advent of nanofabrication, Spin-Hall effect has immense potential for spin-controlled nanophotonic devices [12,39,40]. This study may find direct application in enhancement of Valley-Hall effect [41] and secondary emission [24].

**Funding information.** DST-Nanomission Grant, India (SR/NM/NS-1141/2012(G)) and Center for Energy Science, DST India (SR/NM/TP-13/2016).

**Acknowledgment**. VK acknowledges DST-SERB India for National Post-Doctoral Fellowship reference no. PDF/2016/002037.